\documentclass[11pt]{article}
\usepackage[left=2.5cm,right=2.5cm,top=2.5cm,bottom=2.5cm,a4paper]{geometry} 
\usepackage{graphicx}
\usepackage{amsmath,amssymb}
\usepackage{physics}
\usepackage[affil-it]{authblk}
\newcommand{\gev}{\,{\rm GeV}}

\newcommand{\nnn}{\nonumber\\}
\newcommand{\mc}{\mathcal}
\usepackage{bm}
\usepackage{cite}
\usepackage{comment}

\begin{document}

\title{Five dimensional $SU(5) \times U(1)_{\rm PQ}$ grand unification}

\author{Norimi Yokozaki\thanks{n.yokozaki@gmail.com}}
\author{Junhao Zhu\thanks{junhaozhu@zju.edu.cn}}

\affil{{\small Zhejiang Institute of Modern Physics and Department of Physics, Zhejiang University, Hangzhou, Zhejiang 310027, China}}

\date{}

\maketitle

\begin{abstract}
\noindent

We propose a grand unified model in five dimensions that addresses the strong CP problem. In this framework, the boundary conditions that break the unified gauge group into the Standard Model gauge group simultaneously explain the mass splitting of the new matter content required for gauge coupling unification and the emergence of the QCD axion as the $U(1)_{\rm PQ}$ gauge field. The quality of the axion solution is inherently ensured by a one-form  symmetry of the model. We further investigate the implications of this setup, including its predictions for the proton decay rate, the detectability of axion dark matter, and its cosmological impact. Notably, the model predicts a proton decay signal within the sensitivity of upcoming experiments, a result driven by the trade-off between the unification scale and the reheating temperature. In our scenario, the scales of grand unification and the axion decay constant are naturally unified.

\end{abstract}

\clearpage

\section{Introduction}
For decades, the unnaturalness of certain parameters in the Standard Model (SM) has driven the search for new physics. Three major naturalness problems stand out: the cosmological constant problem, the Higgs hierarchy problem, and the strong CP problem. Each of these involves values that are far smaller than theoretical expectations, including quantum corrections. Achieving these tiny values requires extreme fine-tuning.

Remarkably, two of these issues can be addressed through the anthropic principle. The well-known version of the anthropic principle states that only universes with parameters suitable for creating life can be observed. This selection bias leads to the peculiar values seen in the SM. For the cosmological constant problem, Weinberg \cite{Weinberg:1987dv} derived an upper bound by arguing that life could not exist if the universe expanded too quickly to allow galaxy formation, thereby explaining the smallness of the cosmological constant. For the Higgs mass, the requirement for stable atomic elements beyond hydrogen limits the parameter to a narrow range, with the observed value falling naturally within it \cite{Agrawal:1998xa,Agrawal:1997gf}. Thus, within the anthropic landscape, these two naturalness problems find an explanation. However, the strong CP problem remains an exception.

The strong CP problem  concerns the extremely small parameter $\theta$ in the strong CP-violating term (see e.g.  \cite{Kim:2008hd} for a review). This $\theta$ appears in the following term: 
\begin{eqnarray*}
    \mathcal{L} \ni \frac{\theta}{32 \pi} \, ^\star G^{\mu \nu} G_{\mu \nu},
\end{eqnarray*}
and is related to an observable: the electric dipole moment of neutrons\cite{Crewther:1979pi}.  Current experiment gives a very strict upper bound: $\theta \leq 10^{-10}$ \cite{Abel:2020pzs}. It is argued that a slight increase in $\theta$ by several orders of magnitude would not lead to dramatic consequences in nuclear physics~\cite{Ubaldi:2008nf,Donoghue:2003vs,Banks:2003es}. Moreover, even if one attempts to correlate $\theta$ with other anthropically constrained quantities, achieving such a small value for $\theta$ remains highly challenging~\cite{Carpenter:2009zs,Dine:2018glh}. Therefore, the strong CP problem is still an important naturalness problem to be answered.

By far, the axion solution remains the most promising solution to the strong CP problem~\cite{Peccei:1977hh,Peccei:1977ur,Weinberg:1977ma, Wilczek:1977pj}, being also a dark matter candidate~\cite{Preskill:1982cy,Abbott:1982af, Dine:1982ah}. The idea is to introduce a global $U(1)$ Peccei-Quinn (PQ) symmetry which is broken by the QCD anomaly: 
\begin{eqnarray*}
    \mathcal{L} \ni  -\frac{1}{f_a} \frac{a(x)}{32 \pi} \,^\star G^{\mu \nu} G_{\mu \nu},
\end{eqnarray*}
where $a(x)$ is the Nambu-Goldstone boson of the Peccei-Quinn (PQ) symmetry. If this is the only term that violates the PQ symmetry, the non-perturbative effect of QCD would generate a vacuum energy which is minimized for $\theta - a/f_a=0$, so that the axion field $a(x)$ picks up a CP conserving vacuum. Nonetheless, the existence of other PQ symmetry breaking terms would misalign this mathcing, leading to non-zero strong CP violation. The tininess of $\theta$ is traded for the high quality of the PQ symmetry. However, the quality of the PQ symmetry is easily spoiled by Planck scale physics~\cite{Barr:1992qq, Kamionkowski:1992mf, Holman:1992us, Randall:1992ut} as the quantum gravity is expected to violate global symmetries~\cite{ Svrcek:2006yi,Alonso:2017avz}. For instance, a PQ-breaking term must be suppressed by up to the order of $M_{PL}^{-9}$ for $f_a = 10^{12}$\,GeV, where $M_{PL}$ is the reduced Planck mass.

Another important indication from the SM for new physics is unification. The quantization of the hypercharges of the matter fields suggests the existence of a simpler unified theory~\cite{Georgi:1974sy}. Also this charge arrangement is just right for the SM anomaly cancellation, making it unlikely to be a coincidence. However, grand unified theories (GUTs) face several significant challenges. One major issue arises when embedding the Higgs field into a unified multiplet. After symmetry breaking, the colored triplet component of the multiplet remains and can mediate proton decay. To suppress this process, the triplet must acquire a mass much heavier than the electroweak scale. This large mass hierarchy, known as the doublet-triplet splitting problem, demands a natural explanation beyond fine-tuning.

There have been numerous methods proposed to solve these two seemingly unrelated problems. In particular, there are scenarios that combine GUT gauge symmetries with the global $U(1)_{\rm PQ}$ symmetry in four dimensions~\cite{Wise:1981ry, Kim:1983ia, Davidson:1983fy, Fukuyama:2004zb, Bajc:2005zf, Boucenna:2017fna, Ernst:2018bib, DiLuzio:2018gqe, FileviezPerez:2019fku, DiLuzio:2020qio, Chen:2021haa, Agrawal:2022lsp, Antusch:2023jok, Takahashi:2023vhv}. By considering the extra-dimensional framework, we note that there exists a particularly simple solution to both of them.

A high-quality axion can be realized in a 5D theory, where the fifth dimension is compactified on the orbifold $S^1/Z_2$~\cite{Choi:2003wr}. An additional $U(1)$ gauge field is introduced with $Z_2$ odd parity, ensuring that only the fifth component of the gauge field has a massless Kaluza-Klein (KK) mode, while all other components are suppressed. Due to the residual gauge symmetry, the zero mode of $C_5$ undergoes a symmetry transformation analogous to that of an axion:  
\begin{eqnarray*}  
	C_5 (x) \rightarrow C_5(x) + \alpha .  
\end{eqnarray*}  
By introducing the appropriate 5D Chern-Simons term, $C_5$ can be identified as the QCD axion. 
The quality of the axion is ensured by a one-form symmetry~\cite{Craig:2024dnl} (see also Ref.~\cite{Reece:2023czb}), from the perspective of higher-form symmetries~\cite{Gaiotto:2014kfa}.

The problems of GUT find a particularly simple solution within the 5D framework~\cite{Kawamura:2000ev, Altarelli:2001qj, Hall:2001pg, Hebecker:2001wq} (see also Ref.~\cite{Kawamura:1999nj}), where the fifth dimension is compactified on the orbifold $S^1/Z_2 \times Z_2'$. The unified gauge group $SU(5)$ is broken by assigning different $Z_2 \times Z_2'$ parities to the matter and gauge fields, ensuring that the low-energy effective theory in 4D reduces to the SM. Notably, the $SU(3)$ triplet of the Higgs multiplet does not have a zero mode, naturally explaining its heavy mass hierarchy.

Given these two elegant mechanisms within the 5D framework, it is natural to combine them so that both $SU(5)$ breaking and the axion's anomalous coupling arise from the same underlying mechanism. In this paper, we explore this possibility. As a result, the model predicts a QCD axion that could be detected by future dark matter search experiments, such as DMRadio-GUT. 

To achieve gauge unification, additional $SU(5)$ multiplet fermions are introduced in the bulk and assigned appropriate $Z_2$ parities. We cannot simply introduce arbitrary incomplete multiplets at arbitrary mass scales. The unification scale is constrained by proton decay, and our model notably predicts a proton decay rate within a range detectable by future experiments. This result stems from the trade-off between the reheating temperature and the unification scale. A higher unification scale imposes a stricter upper bound on the reheating temperature. Consequently, to ensure that the reheating temperature remains sufficient to generate the observed baryon asymmetry, the proton decay rate naturally falls within an observable range.

\section{Model}
We consider a $SU(5) \times U(1)_{\rm PQ}$ model in a flat five-dimensional space-time with $S_1/(Z_2 \times Z_2')$ compactification. The physical interval we consider is $y=[0,L=\pi R/2]$, where $R$ is the radius of the fifth dimension. 
The reduced Planck scale $M_{PL}$ is related to the 5D Planck scale $M_{5D}$ as $M_{PL}^2 = M_{5D}^3 L$.
See table~\ref{tab:table} for the assignment of fields.

\begin{table}
    \caption{Location of the fields}
    \begin{center}
        \begin{tabular}{c|c}
            brane ($y=0$) &  SM matter fields, \ $m_{24} (=m_{adj})$ and \ $m_5(=m_2)$ \\
            \hline
            bulk & $H_5$, $\psi_{5,i}$, $\psi_{\bar 5,i}$, $\Sigma_{24}$ and gauge fields \\
        \end{tabular}
    \end{center}
    \label{tab:table}
\end{table}

The gauge group $SU(5)$ is broken into the SM gauge group with boundary conditions~\cite{Kawamura:2000ev, Altarelli:2001qj, Hall:2001pg, Hebecker:2001wq}. The fifth component of the $U(1)_{\rm PQ}$ gauge boson has a zero mode and is identified as an axion~\cite{Choi:2003wr}. 
By defining $P={\rm diag}(1,1,1,1,1)$ and $P'={\rm diag}(-1,-1,-1,1,1)$,
the boundary conditions for the gauge fields of $SU(5)$, $A_M^a=(A_\mu^a, A_5^a)\ (a=1 \dots 24)$, are given by 
\begin{eqnarray}
    Z_2 &:&  A_\mu(y) \to A_\mu(-y) = P A_\mu(y) P,  \ \ 
    A_5(y) \to A_5(-y) = -P A_5(y) P,
    \nnn
    Z_2'&:&  A_\mu(y') \to A_\mu(-y') = P' A_\mu(y') P', \ \   
    A_5(y') \to A_5(-y') = -P' A_5(y') P',
\end{eqnarray}
where $y'=y + L$, and $A_M = A_M^a T^a$ with $T^a$ being generators of $SU(5)$. Only the components, 
$A_\mu^a T^a\,(a=1 \dots 12)$, take the $(+,+)$  ($Z_2$, $Z_2'$) parity and have zero modes. These zero modes correspond to the SM gauge fields; $A_\mu^{a'} T^{a'}\,(a'=13 \dots 24)$ take the $(+,-)$ parity and have masses of $(2n-1)/R\  (n \in \mathbb{N})$. These components correspond to the X/Y gauge boson. (For the KK expansion of the fields, see appendix.)

The Higgs field, $H_5$, contains a $SU(2)_L$ doublet and a $SU(3)_c$ triplet. As the 4D effective theory contains only a $SU(2)_L$ doublet, $H_5$ needs to satisfy
\begin{eqnarray}
    Z_2 &:&  H_5(y) \to H_5(-y) = P H_5(y),  \nnn
    Z_2'&:&  H_5(y') \to H_5(-y') = P' H_5(y').  
\end{eqnarray}
With the boundary conditions, the $SU(2)_L$ doublet has a zero mode and the $SU(3)_c$ triplet has masses of $(2n-1)/R\ (n \in \mathbb{N})$, after KK expansion. 

To solve the strong CP problem, we introduce the $U(1)_{\rm PQ}$ gauge field, $C_M$, that satisfies
\begin{eqnarray}
    && (-,-):\, C_\mu, \  \ \  (+,+):\,  C_5, 
\end{eqnarray}
where
\begin{eqnarray}
    C_M \to C_M + \partial_M \Lambda(x,y), \label{eq:pq_gauge}
\end{eqnarray}
under the gauge transformation. Here, $C_5$ has a zero mode and is identified as the QCD axion~\cite{Choi:2003wr}:
\begin{eqnarray}
    a(x)/f = \frac{1}{2}\oint C_5 dy.
\end{eqnarray}
The periodicity is assured for $\Lambda(y) = 2 y/R$, which is $(-,-)$ for $Z_2$ and $Z_2'$ parities up to ${\rm mod}\ 2\pi$.

The action is given by~\footnote{To be consistent with the parities, $(1/2)[\delta(y) \pm \delta(y - 2 L)]$ and $(1/2)[\delta(y-L) \pm \delta(y + L)]$, are implicitly understood.}
\begin{eqnarray}
    S = \int d^4 x dy \left[\mathcal{L}_1 \delta(y) + \mathcal{L}_2 \delta(y-L) + \mathcal{L}_{\rm bulk}  \right], 
\end{eqnarray}
where
\begin{eqnarray}
    \mathcal{L}_{bulk} &=& -\frac{1}{2 g_5^2} \Tr(F_{M N} F^{M N} ) 
    -\frac{1}{4 g_{P}^2} C_{M N} C^{M N} 
    \nnn &+& \mc{L}_{CS} 
    + \mc{L}_{Higgs} .
\end{eqnarray}
Here, 
$F_{M N}$ and $C_{M N}$ are the field strength of the $SU(5)$ and $U(1)_{\rm PQ}$ gauge fields, respectively; $g_5$ and $g_P$ are 5D gauge couplings of mass dimension $-1/2$.
$\mathcal{L}_{CS}$ is a Chern-Simons coupling:
\begin{eqnarray}
    \mathcal{L}_{CS} = \frac{\kappa}{64\pi^2} \epsilon^{MNPQR} C_M \tr( F_{NP} F_{QR}),
\end{eqnarray}
where $\kappa \in \mathbb{Z}$ so that the partition function is gauge invariant.

After expanding the gauge fields in their KK modes and integrating $y$ out, we obtain the 4D effective Lagrangian. By canonically normalizing the axion field, the decay constant is identified as
\begin{eqnarray}
    f_a^{-1} = \kappa g_P L^{1/2} =  \kappa g_{P,4D} L,
\end{eqnarray}
where $g_{P,4D}$ is a coupling constant of four-dimensional effective theory.

With the matter content of the SM, the gauge couplings are not unified, and it is difficult to be consistent with our framework. 
In order to be consistent with $SU(5)$, we may need some light fields that are incomplete multiplets of $SU(5)$. Also, the scenario should satisfy the proton decay constraint, $p \to e^+ \pi^0$, which gives $M_X > \mathcal{O}(10^{15})$\gev. Here, $M_X$ is the mass of the GUT gauge boson. The lightest mode has a mass of $R^{-1}$.

\subsection{Gauge coupling unification}

If there are no significant threshold corrections or GUT-breaking brane interactions,\footnote{We may introduce (small) brane kinetic terms, for instance, 
    \begin{eqnarray}
        \mathcal{L}_k &=& \left(
        -\frac{1}{4 \kappa_1^2} (B_{\mu \nu} B^{\mu \nu})
        - \frac{1}{2 \kappa_2^2} \Tr (W_{\mu \nu} W^{\mu \nu})
        - \frac{1}{2 \kappa_3^2} \Tr (G_{\mu \nu} G^{\mu \nu}) 
        \right) \nonumber \\ 
        &\times& \delta(y-L), 
    \end{eqnarray}
which helps unification.} consistency with the $SU(5)$ grand unification requires the presence of new fields at intermediate scales to ensure that the three gauge couplings of the SM (approximately) converge at a single point. 

Since $SU(5)$-complete multiplets contribute equally to the beta functions of the SM gauge couplings, the new fields must be incomplete multiplets of $SU(5)$. Achieving gauge coupling unification is not trivial since we need to introduce matter fields as complete multiplets of $SU(5)$, originally, and then assign $Z_2'$ parity using the operator $P'$. We cannot simply introduce arbitrary incomplete multiplets at arbitrary mass scales.

Fortunately, these fields can be included in a way consistent with our setup, specifically respecting the $Z_2'$ parity expressed by $P'$.

We include the following new fields in the bulk, which transform under $Z_2'$ parity as
\begin{eqnarray}
    \Sigma_{24}(y') &\to& \Sigma_{24}(-y') = P' \Sigma_{24}(y') P' \, ,\\
    \psi_{5,i} (y') &\to& \psi_5 (-y') = P' \psi_{5,i}(y') \, , \\
    \psi'_{\bar 5, i} (y') &\to& \psi'_{\bar 5, i} (-y') = P' \psi'_{\bar 5, i}(y') ,
\end{eqnarray}
where $i=1,2$. Their mass terms are put on the $y = L$ brane. The right-handed fermions do not have zero modes. Due to the boundary conditions, the only zero modes are the $SU(2)_L$ triplet, $SU(3)_c$ octet of $\Sigma_{24}$, and $SU(2)_L$ doublets of $\psi_{5,i}$ and $\psi_{\bar 5,i}$. (The singlet is also contained in $\Sigma_{24}$, but it does not contribute to the RGE running.) 
For simplicity, we assume there is no mixing between the new doublets and the SM fields.\footnote{As the rank of the mass matrix of the doublet fermions remains the same with or without the mixing, it does not necessarily forbid the mixing, as long as all mass parameters are of the same order. In this case, the massless components in the mass eigenbasis can simply be identified as the SM leptons.} 	
The low-energy effective Lagrangian contains the mass terms of these zero-modes:
\begin{eqnarray}
-\mathcal{L}_{\rm eff} &\ni&	\left[ m_{\rm adj}
\left(
 \tr(\overline{\Sigma_{3,L}^c} \Sigma_{3,L}) + \tr(\overline{\Sigma_{8,L}^c} \Sigma_{8,L}) \right)
    + m_2 \overline{ \psi_{2,L_i}^c} \psi'_{2,L_i} \right] + h.c.,
\end{eqnarray}
where $\Sigma_{3,L}$, $\Sigma_{8,L}$, and $\psi'_{2,L_i}$ represent the $SU(2)_L$ triplet, the $SU(3)_c$ octet, and the $SU(2)_L$ doublet, respectively.
With these new fields that are imcomplete multiplets of $SU(5)$, the one-loop renormalization group equations (RGEs) are changed and given by
\begin{eqnarray}
    \frac{d g_1}{d t} &=& \frac{1}{16\pi^2} \left(\frac{41}{10} + \frac{3}{5} \Delta \tilde{h} \right) g_1^3 \\
    \frac{d g_2}{d t} &=& \frac{1}{16\pi^2} \left(-\frac{19}{6} + \Delta \tilde{h}+ \Delta_3 \right) g_2^3 \\
    \frac{d g_3}{d t} &=& \frac{1}{16\pi^2} \left(-7 +  \Delta_8 \right) g_3^3 ,
\end{eqnarray}
where $g_1$ takes the $SU(5)$ normalization: $g_1=\frac{5}{3} g_Y$; $\Delta \tilde{h} = 4/3$, $\Delta_3 = 4/3$ and $\Delta_8 = 2$.

\begin{figure}
    \begin{center}
    \includegraphics[width=0.49\textwidth]{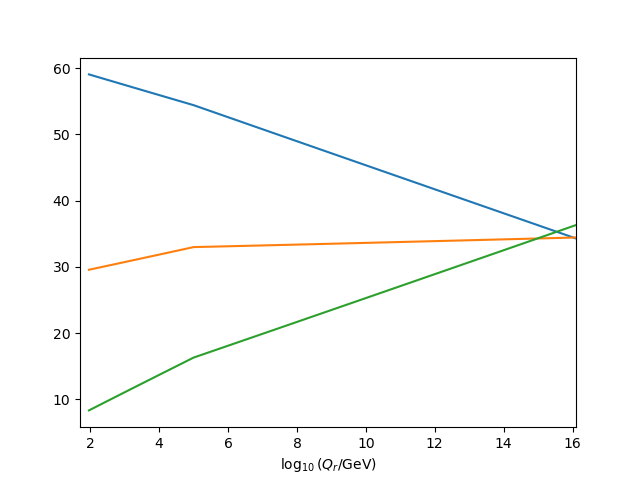}
    \includegraphics[width=0.49\textwidth]{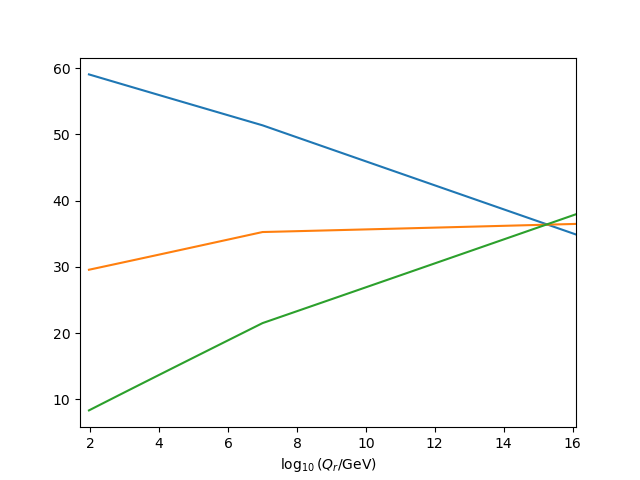}			
    \includegraphics[width=0.49\textwidth]{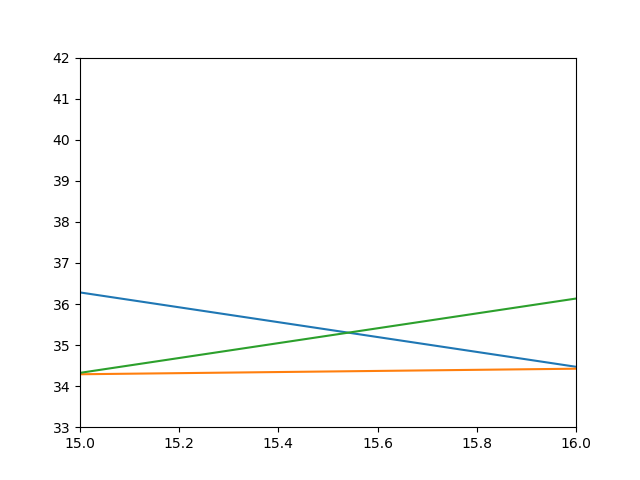}
    \includegraphics[width=0.49\textwidth]{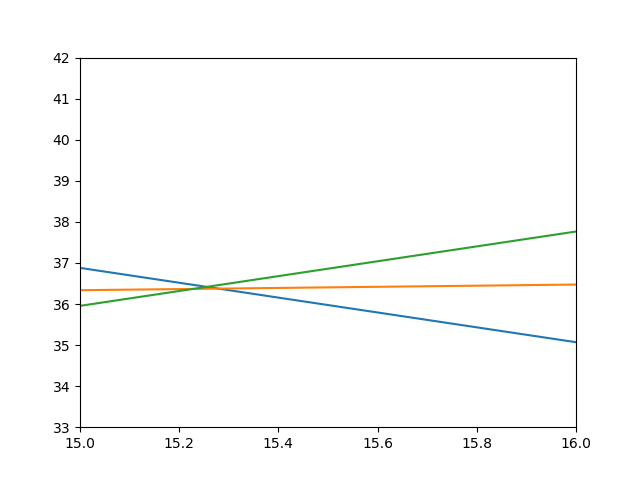}					
    \end{center}
    \caption{The renormalization group running of gauge coupling constants, $\alpha_{1,2,3}^{-1}$ at the two-loop level; $\alpha_1^{-1}$, $\alpha_2^{-1}$ and $\alpha_3^{-1}$ are shown as the top to bottom lines. 
    The horizontal axis shows the renormalization scale, $\log_{10}(Q_r/{\rm GeV})$. The mass for the adjoint fermion is taken as $m_{\rm adj}=10^5\,{\rm GeV}$ for the left panel, 
    $m_{\rm adj}=10^7\,{\rm GeV}$ for the right panel. We take $m_2=m_{\rm adj}$.}
    \label{fig:rge1}
\end{figure}

\begin{figure}
    \begin{center}
        \includegraphics[width=0.49\textwidth]{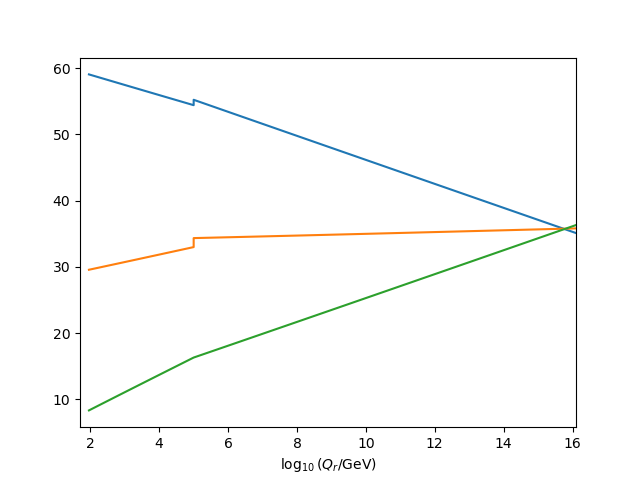}
        \includegraphics[width=0.49\textwidth]{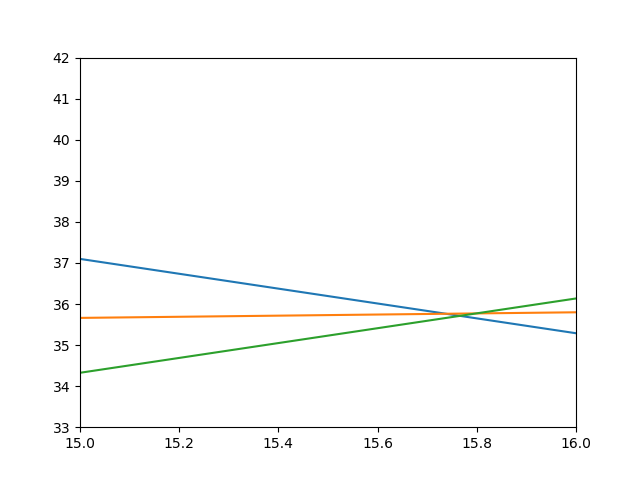}			
    \end{center}
    \caption{The same figure as fig.\ref{fig:rge1} except for $m_2=500 \, m_{\rm adj}$ with 
        $m_{\rm adj}=10^5\,{\rm GeV}$. The effects from the mass difference are included as one-loop threshold corrections.}
    \label{fig:rge2}
\end{figure}

In figures~\ref{fig:rge1} and \ref{fig:rge2}, we plot the running of the gauge coupling constants ($\alpha_{1,2,3}^{-1}$).  For computation, we use two-loop RGEs including the gauge couplings and the top-Yukawa coupling. 
The RGEs are obtained using {\tt PyR@TE 3} package~\cite{Sartore:2020gou}.
We take $m_2 = m_{\rm adj}$ in both panels in figure~\ref{fig:rge1}. In figure~\ref{fig:rge2}, we set $m_{\rm adj} = 10^5 \, \text{GeV}$ and $m_2 = 500 \, m_{\rm adj}$. It can be observed that gauge coupling unification is well achieved by introducing $SU(2)_L$ doublets and adjoint fermions at the intermediate scale. Notably, a larger adjoint mass scale results in a lower unification scale, which, in turn, leads to rapid proton decay via dimension-six operators. On the other hand, a smaller adjoint mass scale is constrained by the successful predictions of standard cosmology, as discussed below. Therefore, only a limited mass range is consistent with the constraints from cosmology and proton decay experiments.

It should be noted that $\Sigma_8$ could decay to three-light quarks with a long life-time. The relevant operator is, for instance,
\begin{eqnarray}
    M_*^{-2} \epsilon^{\alpha \beta \gamma} \overline{d_R}_\delta (\Sigma_{8,L})^\delta_{\ \alpha} \overline{u_R}_\beta (d_R^c)_\gamma,
\end{eqnarray}
where $M_*$ is a cut-off scale, which may be around $M_{5D}$. The life-time is estimated as
\begin{eqnarray}
    \tau \sim 64\pi^3 M_*^4 m_8^{-5} \simeq 1.3 \times 10^{17}\, s \left(\frac{M_*}{10^{17}\,{\rm GeV}}\right)^4 \left(\frac{m_8}{10^6\,{\rm GeV}}\right)^{-5}\ .
\end{eqnarray}
When the temperature larger than $m_8 (=m_{\rm adj})$, the octet fermions are thermalized and its number density over the entropy density is $\approx 4 \times 10^{-3}$. As the remanent of quasi-stable colored particles affects the predictions of Big Bang Nucleosynthesis (BBN) and the cosmic microwave background (CMB) spectrum, the energy density of the quasi-stable color octet fermion is severely constrained~\cite{Kawasaki:2004qu, Kawasaki:2017bqm, Gross:2018zha}. 
To avoid the constraints, we require the reheating temperature, $T_R$, to be smaller than $m_{\rm adj}$ so that the octet fermions could never be created.

\subsection{Proton decay}

Since the expected GUT scale is $\mathcal{O}(10^{15})$ GeV, the constraint from the proton decay ($p \to e^+ \pi^0$) is significant. The dimension six effective operators for the proton decay are
\begin{eqnarray}
    \mathcal{L}_{\rm eff} = \frac{g_{5,4D}^2}{M_X^2} \epsilon_{\alpha \beta \gamma} ( A^{(1)} \overline{u^c}^\alpha P_R d^\beta \overline{e^c} P_L u^\gamma
    + 
    A^{(2)} \overline{u^c}^\alpha P_L d^\beta \overline{e^c} P_R u^\gamma 
    ) + h.c.,
\end{eqnarray}
where $g_{5,4D}$ is the gauge coupling constant of $SU(5)$ in the 4D effective theory, and $M_X = R^{-1}$ is the mass of the GUT gauge boson. The renormalization factors of short distance effects, $A^{(1)}$ and $A^{(2)}$, are written as
\begin{eqnarray}
    A^{(1,2)}=A_{a}^{(1,2)} A_{b}^{(1,2)} A_{c}^{(1,2)},
\end{eqnarray} 
where $A_{a}^{(1,2)}$ is a renormalization factor from $m_Z$ to $m_p$ (proton mass), $A_{b}^{(1,2)}$ is that from the mass scale of the adjoint fermions to $m_Z$, and $A_{c}^{(1,2)}$ is that from the GUT scale to the mass scale of the adjoint fermions. 

At the one-loop level, the renormalization factors are written as~\cite{Buras:1977yy, Wilczek:1979hc}
\begin{eqnarray}
    A_a^{(1,2)} = \prod_{i=1}^3 \left( \frac{\alpha_i(m_p)}{\alpha_i(m_z)} \right)^{\frac{\tilde{\gamma}_i^{(1,2)}}{b_i}},
\end{eqnarray}
where $b_i$ is the coefficient of the one-loop beta function, $\tilde{\gamma}_i^{(1)} = (-2,-9/4, -23/12)$ and $\tilde{\gamma}_i^{(2)} = (-2,-9/4, -23/12)$. The other renormalization factors are obtained by the replacements:
$A_b^{(1,2)} = A_a^{(1,2)} \{m_z \to m_8, m_p \to m_z, b_i \to b_i + \Delta b_i \}$ and $A_c^{(1,2)} = A_a^{(1,2)} \{m_z \to M_X, m_p \to m_8, b_i \to b_i + \Delta' b_i\}$.
We use $A^{(1)}=3.3$ and $A^{(2)}=3.7$.  

Together with the Wilson coefficients, the proton decay rate is estimated using matrix elements calculated by lattice~\cite{Yoo:2021gql}:
\begin{eqnarray}
    \Gamma(p \to e^+ \pi^0) \simeq \frac{m_p}{32\pi}
    \left[1-\left(\frac{m_{\pi}}{m_p}\right)^2
    \right]
    \sum_I \left|C^I W_0^I(p \to \pi^0) \right|^2, \label{eq:protondecay}
\end{eqnarray}
where $|C^I| = A^{(1,2)} g_{5,4D}^2/M_X^2$ and $W_0^I \approx 0.11 {\,\rm GeV}^2$ are the Wilson coefficient and the matrix element, respectively, and $I$ represents the type of operators. Here, we have neglected the electron mass.~\footnote{The interference term is proportional to the electron mass.}

The experimental constraint from Super-Kamiokande is $\Gamma^{-1} > 2.4 \times 10^{34}\,{\rm years}$~\cite{Super-Kamiokande:2020wjk}. By using Eq.~\eqref{eq:protondecay}, the mass of the GUT gauge boson should satisfy
\begin{eqnarray}
    M_X \gtrsim 4.5 \times 10^{15}\, {\rm GeV}.
\end{eqnarray}
Requiring that the gauge couplings of the SM approximately unify at a scale larger than \( M_X \), the mass of the color octet fermion must be smaller than \( 10^5 \mathchar`- 10^6 \, {\rm GeV} \) (see Fig.~\ref{fig:rge1},\ref{fig:rge2}). To ensure that the octet fermions are never produced, we impose \( T_R < m_8 \). Consequently, the reheating temperature is bounded from above as
\begin{eqnarray}
    T_R \leq 10^5 \mathchar`- 10^6 \, {\rm GeV} . \label{eq:trlimit}
\end{eqnarray}
With such a relatively low reheating temperature, thermal leptogenesis~\cite{Fukugita:1986hr} cannot successfully account for the observed baryon asymmetry of the universe~\cite{Giudice:2003jh}. 
However, it is possible that the right-handed neutrinos are directly produced from inflaton decay, and their subsequent decay generates sufficient \( B-L \) asymmetry to explain the observed baryon asymmetry~\cite{Lazarides:1990huy, Giudice:1999fb, Asaka:1999yd, Asaka:2002zu}. It is also noted that when the mass splitting among the right-handed neutrinos is small, there is an enhancement of the asymmetry~\cite{Pilaftsis:1997jf, Pilaftsis:2003gt}.
In these scenarios, the reheating temperature can be low and remain consistent with Eq.~\eqref{eq:trlimit}.

Note that the proton decay induced by the exchange of the colored Higgs is not significant. Unlike supersymmetric $SU(5)$, the proton decay $p \to K^+ \overline{\nu}$ does not give important constraints.

\subsection{Axion cosmology and detection}

The axion in our model is a candidate for dark matter. The axion decay constant $f_a$ is $\sim R^{-1}$, and the axion mass is estimated as~\cite{GrillidiCortona:2015jxo}
\begin{eqnarray}
    m_a \approx 10^{-9} {\rm eV} 
    \left(\frac{5 \times 10^{15}\, {\rm GeV} }{f_a}\right)
\end{eqnarray}

The coherent oscillation of the axion field can explain the present energy density of dark matter, depending on the initial value. 
We consider the case where the PQ symmetry is broken before or during the inflation, and is never restored afterward. 
The axion field starts to oscillate when the Hubble parameter becomes close to the axion mass, $m_a(T)$. Using the approximated power law of the axion mass with respect to the temperature, $m_a(T) \propto T^{-4}$, the energy density is estimated as~\cite{Borsanyi:2016ksw}    
\begin{eqnarray}
    \Omega_{\rm axion} h^2 \approx  0.2 \times 10^{4} \theta_0^2 
    \left( \frac{f_a}{5 \times 10^{15}\,{\rm GeV} } \right)^{7/6} \,.
\end{eqnarray}
By assuming the axion explains all the energy density of dark matter, the initial constant value of the axion field, $\theta_0 \equiv a(x)/f_a$, needs to be $\mathcal{O}(10^{-2})$.

During inflation, the axion field may acquire large fluctuations when the inflation energy is high. These fluctuations are inhomogeneous and contribute to the energy density of the axion in addition to the homogeneous contribution. 
The observed CMB power spectrum by the Planck satellite~\cite{Planck:2019nip} is consistent without the inhomogeneous contribution; therefore, the Hubble scale during inflation, $H_I$, is constrained as~\cite{Hertzberg:2008wr, OHare:2024nmr}
\begin{eqnarray}
    H_I \lesssim   1.4 \times 10^{9}\,{\rm GeV}
\left(\frac{\theta_0}{10^{-2}}\right)	
    \left(\frac{f_a}{ 5 \times 10^{15}\,{\rm GeV}}
    \right).
\end{eqnarray}

Lastly, we discuss the possible detection of the axion. 
Since our model is based on the grand unified theory, the predicted ratio of the electromagnetic coupling to the QCD coupling is $E/N=8/3$, and the decay constant $f_a$ is of the order of the GUT scale, $R^{-1} \sim L^{-1}$. Taking into account the mixing with the neutral pion, the axion photon coupling is estimated as~\cite{GrillidiCortona:2015jxo}
\begin{eqnarray}
    g_{a \gamma \gamma} \approx   \frac{\alpha}{2\pi f_a}(8/3 - 1.92)
    \approx 
    1.7 \times 10^{-19}\,{\rm GeV}^{-1} 
    \left(
    \frac{5 \times 10^{15}\,{\rm GeV} }{f_a}
    \right),
\end{eqnarray}
where the coupling is defined as
\begin{eqnarray}
    \mathcal{L}_{eff} \ni \frac{1}{4} g_{a \gamma \gamma} a(x) \left(   
    \frac{\epsilon^{\mu\nu\rho \sigma}}{2} F_{\mu \nu} F_{\rho \sigma} 
    \right). 
\end{eqnarray}
Provided that the QCD axion in our model accounts for the entire relic abundance of dark matter, future dark matter search experiments targeting axion dark matter, particularly DMRadio-GUT, may detect this QCD axion~\cite{ABRACADABRA:2018rtf,Salemi:2021gck,DMRadio:2022jfv}.

\section{Conclusion}

We have considered a $SU(5) \times U(1)_{\rm PQ}$ grand unified model in 5D space-time with the extra dimension compactified on $S^1 /(Z_2 \times Z_2')$. The $SU(5)$ group is broken into the standard model gauge group by the boundary conditions. The zero mode of the fifth component of the $U(1)_{\rm PQ}$ gauge field is identified as a QCD axion. The quality of the axion solution is guaranteed by the one-form symmetry~\cite{Craig:2024dnl}. 

The axion in our model is a dark matter candidate. Its mass is estimated as $\order{10^{-9}} {\rm eV}$, and the axion photon coupling is estimated as $1.7 \times 10^{-19}  {\rm GeV}$ . Provided the axion in our model accounts for the entire relic abundance of the dark matter, it could be detected in future axion dark matter search experiments.

New matter contents are introduced to achieve gauge coupling unification, including a \( SU(2)_L \) triplet, a \( SU(3)_c \) octet, and two \( SU(2)_L \) doublet fermions. These particles arise from \( SU(5) \) multiplets, and the required mass splitting is consistently explained through the \( Z_2 \times Z_2' \) parity used for the GUT group breaking.
The unification scale is expected to be $\mathcal{O}(10^{15}) \, {\rm GeV}$, which faces severe constraints from the proton decay channel $p \to e^+ \pi^0$.

Requiring that the colored octet does not disrupt the successful predictions of standard cosmology imposes a trade-off between the reheating temperature and the unification scale. A larger unification scale results in a stricter upper bound on the reheating temperature. Therefore, to ensure that the reheating temperature is not too low to generate the observed baryon asymmetry, the proton decay rate is expected to fall within an observable range. This proton decay rate is within a range that can be easily detected in future experiments.

Since the reheating temperature must be smaller than the mass of the octet fermion, it is bounded above at \( 10^5 \mathchar`- 10^6 \, {\rm GeV} \). To account for the baryon asymmetry of the universe, resonant and/or non-thermal leptogenesis may need to be considered.

\section*{Acknowledgments}
N. Y. thanks R. Kitano for discussions.
N. Y. is supported by a start-up grant from Zhejiang University. 
J. Z. is supported by Zhejiang University.

\appendix

\section{KK expansion} \label{sec:kk}

The expansions of bulk fields without masses are summarised as 
\begin{eqnarray}
    \Phi_{++}(x,y) &=& \sqrt{\frac{2}{L}} \sum_{n=0}^\infty  \frac{1}{\sqrt{2^{\delta_{n 0}}   }}\cos( \frac{2 n }{R} y) \phi_{++}^n(x), \nnn
    \Phi_{--}(x,y) &=& \sqrt{\frac{2}{L}} \sum_{n=0}^\infty \sin( \frac{(2n+2) }{R} y) \phi_{--}^n(x), \nnn
    \Phi_{+-}(x,y) &=& \sqrt{\frac{2}{L}} \sum_{n=0}^\infty  \cos(  \frac{(2n+1)}{R} y) \phi_{+-}^n(x), \nnn
    \Phi_{-+}(x,y) &=& \sqrt{\frac{2}{L}} \sum_{n=0}^\infty  \sin(  \frac{(2n+1)}{R} y) \phi_{-+}^n(x),
\end{eqnarray}
where only the field with $(+,+)$ boundary condition have zero mode.

\providecommand{\href}[2]{#2}\begingroup\raggedright\endgroup

\end{document}